# Integrating Superconductive and Optical Circuits

F. Stella [1], M. Casalboni [2], M. Cirillo [2,3], V. Merlo [2,3], C. Palazzesi [2], G.P. Pepe [4], P. Prosposito [2,3], and M. Salvato [2,4]

1) Dipartimento di Ingegneria Meccanica, Università di Roma *Tor Vergata*, 00133 Roma, Italy

2) Dipartimento di Fisica and MINAS-Lab, Università di Roma *Tor Vergata*, 00133 Roma, Italy

3) CNR-INFM *Coherentia* and Università di Napoli *Federico II*, 80126 Napoli, Italy

4) SuperMat-INFM, Università di Salerno, 84081 Baronissi , Italy

We have integrated on oxidized silicon wafers superconductive films and Josephson junctions along with sol-gel optical channel waveguides. The fabrication process is carried out in two steps that result to be solid and non-invasive. It is demonstrated that 660 nm light, coupled from an optical fibre into the channel sol-gel waveguide, can be directed toward superconducting tunnel junctions whose current-voltage characteristics are affected by the presence of the radiation. The dependence of the change in the superconducting energy gap under optical pumping is discussed in terms of a non-equilibrium superconductivity model.

The interaction of light with superconducting tunnel structures has received in the past great attention both at theoretical and experimental level. Studies were carried out for investigating nonequilibrium properties of the superconducting state [1], photo-sensitive CdS semiconductor barriers [2], nonlinear dynamics and phase-locking [3] as well as ultrafast device physics [4]. A main constraint for all experiments was always the coupling between the light, generated at room temperature, and the Superconductive Tunnel Junctions (STJs) or films operating at cryogenic temperatures. Usually, cryostats with optical windows are employed for direct illumination, but this approach often requires sophysticated mechanical set up, demonstrated to be very useful in many fundamental physics investigations, but quite inadequate for device physics applications. Problems of focusing are the main limitation to a direct illumination from a fibre inserted into the cryogenic inset hosting the chip, and the coupling between light and the superconducting films/tunnel junctions is often challenging in this case. In this paper we report on a possible solution of this problem based on merging the integration of two technologies, namely the sol-gel techniques for channel waveguides fabrication [5] and the trilayer technology for Nb-AlOx-Nb STJs [6].

In a previous publication it was already demonstrated that the niobium trilayer technology is fully compatible with sol-gel techniques and methods [7]. However, the fabrication procedure herein presented is different since the steps are now "inverted" and 2-D waveguide definition has been applied to the planar film; in particular, we first grow the typical Nb-NbAlOx-Nb trilayer on oxidized silicon wafers and only once the junction fabrication is fully accomplished, we deposit all over the wafer 1μm of sol-gel material. The geometry of the sol-gel is then defined by optical lithography and ion-gun etching ; the final result of our fabrication process is sketched in Fig. 1, where the coupling of the light coming from the optical fibre to the channel guide and then to the STJs is clearly shown. In Fig. 1a, apart for the sol-gel strips indicated by the arrows, the pattern is relative to the all Nb trilayer. The localization of the STJs in Fig. 1 is indicated by the black-filled areas.

A SEM photograph of a fabricated chip is shown in Fig. 2a: a square (10x10 µm$^2$) window junction is clearly visible, and very close to it (on the right) a 4 µm-wide sol-gel channel waveguide. The overall process resulted to be very stable : the STJs had good quality current-voltage characteristics, and the thermal cycling from 300K to 4.2K did not affect the mechanical properties of the realized structures.

The hybrid organic/inorganic planar optical waveguides were synthesized by means of sol-gel technique. The starting precursors were Zr(IV)-Propoxide (Zr) and 3-Glycidoxy propyltrimethoxisilane (Glymo); molar ratios of solvents and precursors used for the synthesis of the samples were Zr:Glymo:water:2-methoxyethanol =1:1:0, 1:0, 4. Silicon substrates were chosen with a 8 µm thick thermally grown silicon oxide layer in order to provide the adequate optical insulation from the high refractive index of silicon. The liquid solution was first stirred for few minutes at room temperature, then it was filtered and deposited by spin coating on the substrate. Films have been cured at 120 °C with ramp up and ramp down procedure. A pattern consisting of 4µm width channels was transferred to the planar films by means of 1.5 µm thick photoresist layer. The subsequent lateral definition of the 2-D optical confinement was obtained by means of an ion beam etching process. The thickness of the final channel structures was measured using an α-step profilometer and it was determined to be of the order of 1 µm. A SEM image of a 2-D waveguide is reported in the inset of figure 2 b, showing the good lateral definition of the confined structure.

The refractive index measurements of planar films of Zr/Glymo were reported in a previous work [7]. The relative spectral losses for our planar films, measured by scattering detection technique were about 3,6 dB/cm at 514 nm, as shown in Fig.2b. This technique is based on the direct monitoring of the scattered light intensity out of the plane of the guide using a digital camera. The acquired pictures are analysed by means of a specific software which converts the light intensity in a grey scale. The propagation loss coefficient can then be determined by mapping the decay of scattered light intensity along the propagation length of the guide using the following expression $y=y_0+A\exp(-\alpha x)$ where $y_0$ represents the constant background noise (in terms of light intensity), x the travelled distance (expressed in cm) and α is the extinction coefficient (expressed in cm$^{-1}$). The optical losses of 2-D structures (4 μm wide) have been measured by means of the same technique and they resulted a bit higher; this is probably related to the lateral waveguide definition process and it has to be further investigated. The light generated by a 660 nm source was brought from an optical fibre down to the chip holder of the cryostat. The light was coupled, at room temperature, into the channel waveguide by the help of a mechanical manipulator and optical microscope : the losses when stepping from the optical fibre into the channel waveguide brings the light power down of two orders of magnitude. This measurement was performed at room temperature by measuring, for a test channel waveguide, the intensity at the input and at the output of the waveguide. Accordingly, we estimated a power of few microwatts was reaching the junction sites. Moreover, we performed a calibration of the laser light power as a function of the nominal current output. It is worth noting that the calibration of the output power as a function of the laser current, performed at the transition between fibre and channel waveguide gave us a linear dependence when the laser current was above the value of 50mA.

In Fig. 3a the effect of the light irradiation on a portion of the subgap current-voltage (I-V) characteristic of a square 10 μm-side junction is presented (the complete I-V curve of the inset accounts for the good quality of fabricated STJs). An increase of the subgap current up to 200nA is clearly observed in correspondence of only few microwatt of incident power. The change in the gap voltage $\delta V_g$ of the 10μm x 10μm junction induced by the light irradiation has been measured at fixed bias currents by detecting the the voltage change by means of a a lock-in amplifier synchronized at the light chopper frequency. Results are presented in Fig. 3b. The linearity of $\delta V_g$ vs the incident laser power can be analyzed in terms of a non-equilibrium change of the excess quasiparticles (qp) number *n* after the steady-state illumination according to the stationary solution of the Rothwarf-Taylor equations [8]. In the limit of weak optical pumping, n can be written as $n \approx \dfrac{I_{qp}\tau_R}{8N(0)\Delta_0}$, where $I_{qp}$ is the qp optical pumping rate, $\Delta_0$ is the unperturbed energy gap, $\tau_R$ the effective qp recombination time, and N(0) the density of states at the Fermi level. We assume for Nb: $N(0) = 31.7 \times 10^{21}$ states/(eVcm$^3$), $\Delta_0 = 1.35$meV as measured from I-V curves, $\tau_R = 0.35$ns at T=4.2K according to Ref. 9. The optical pumping rate can be expressed in the form $I_{qp} = W/d\Delta$ where d is the thickness of the absorbing electrode (d=440nm), and W is the laser power density (expressed in watt/cm$^2$). In the limit of a weak optical perturbation, the excess qp concentration $n$ is demonstrated to be proportional to the normalized energy gap change [10], i.e. $\frac{-\delta\Delta}{\Delta_0} = 2n \approx 2 \times 10^{-4} W$.

Experimental results presented in Fig. 3b at a bias current $I_b=50\mu A$ corresponding to the sumgap voltage state and at a laser current of 90mA gives $\delta V_g/\Delta_0 = 0.0016$, and hence a laser power density of $W \sim 8$ watt/cm$^2$. The vertical error bar for all these data is few nV. If we assume a uniform illumination of the top electrode surface in front of the wave-guide output –this assumption is reasonable due to the different heigts between waveguide and the top plane of the junction-, the output power of the fibre results ~3 μwatt, i.e. a value quite consistent with the room temperature overall dispersion measurements across the planar wave-guide as previously discussed. This demonstrates the consistency of the nonequilibrium picture as assumed in data analysis, and it opens the way also to exciting future investigations about the role of nonequilibrum on Josephson devices through an extremely high control of the optical pumping.

In conclusion, we have reported on the merging of two integrated technologies, namely the all niobium trilayer superconductive process and the sol-gel technique for the fabrication of optical channel waveguides. The results that we have obtained give consistent account for the effect of nonequilibrium quasi-particle dynamics in tunnel juctions and lead us to the conclusion that the interaction of light and light pulses with superconducting structures can be quantitatively handled by adequate design and development of the techniques that we have herein presented.

We thank professor Maria Letizia Terranova and coworkers for the the SEM analysis and pictures. We also ackowledge the precious help of Dr. Ivano Ottaviani in sample fabrication, and Dr Vito Pagliarulo in the assembly of cryogenic optics.

**Figure Captions**

Figure 1 : Schematic top view (a) and cross section (b) , both not to scale, showing the final result of our fabrication process. We first define the trilayer structure (see top view) with relative tunnel junctions (black-filled areas) then we deposited above the trilayer the sol-gel layer whose geometry is defined by protective resist and ion milling (the patterned areas are indicated by the arrows). The light is brought to the junctions by the channel waveguide collecting the signal from the optical fibre.

Figure 2 : (a) SEM picture of one 10μm-side junction with a channel waveguide coupled to it; (b) A typical behaviour of the optical losses for a planar film of Zr/Glymo; the inset is a SEM image (top view) showing the good lateral definition of a 6μm-wide waveguide.

Figure 3 : (a) Light on (top) - light off (bottom) picture of a portion of subgap curve of the current-voltage characteristic of a 10 μm Josephson junction. The inset shows the current-voltage characteristic of the junction (horizontal 1mV/div, vertical 100 μA/div ; (b) Dependencies of the voltage shifts generated by 660 nm-light irradiation for different values of the dc bias current. All the data in this figure were obtained at 4.2 K.

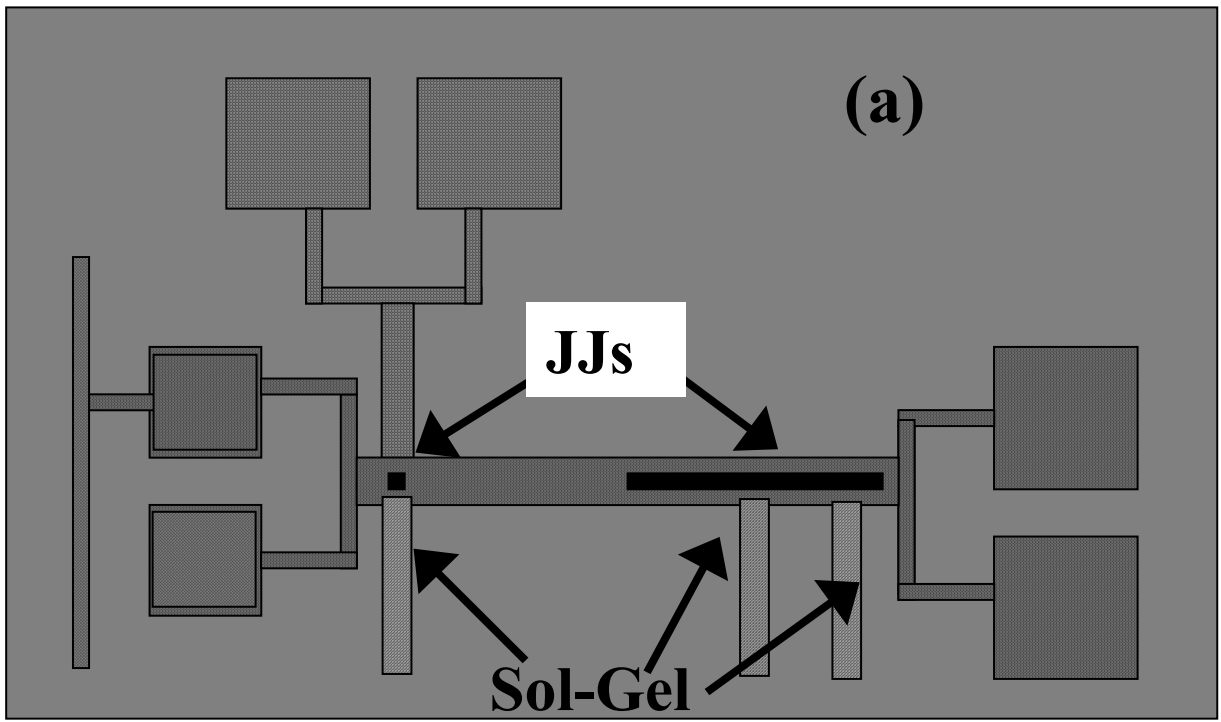

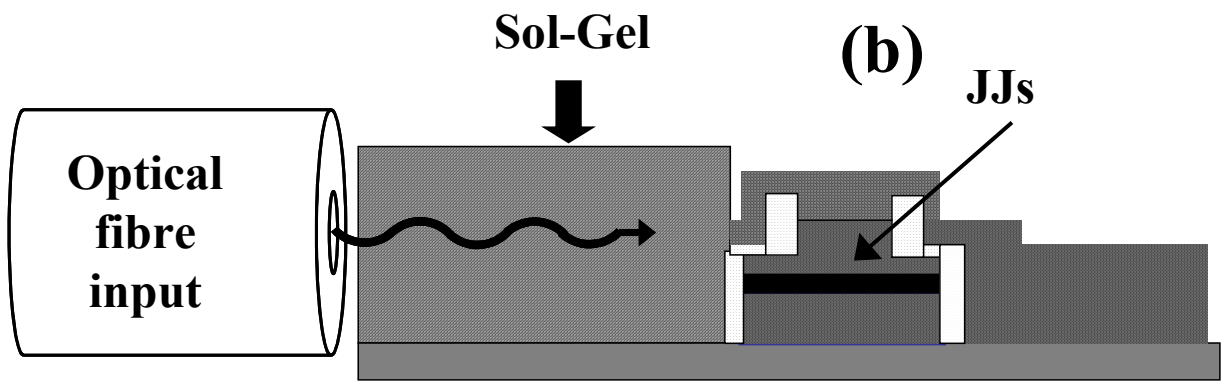

F.Stella et al., Fig. 1

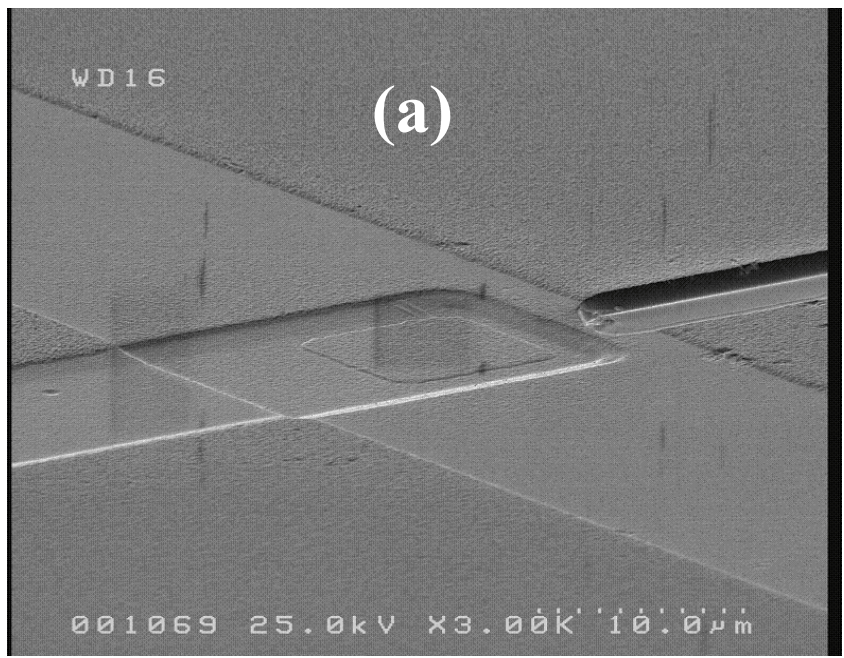

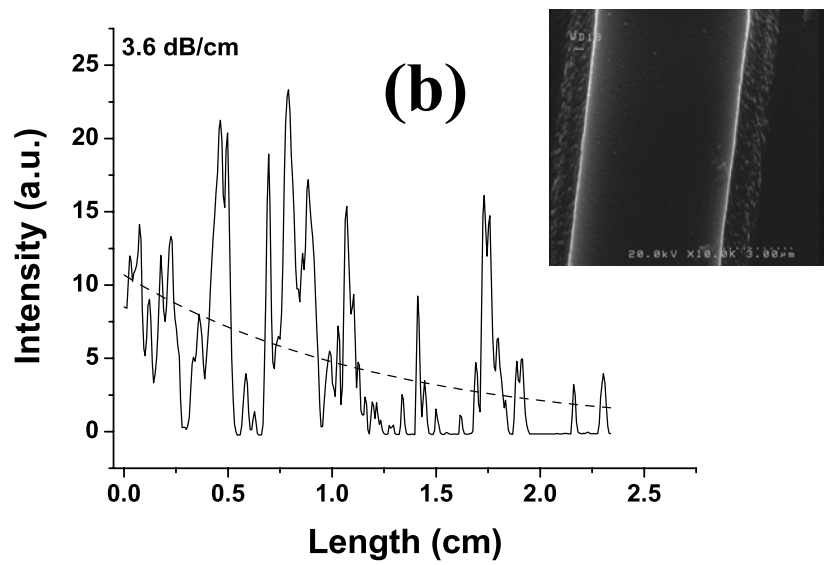

F.Stella et al., Fig. 2

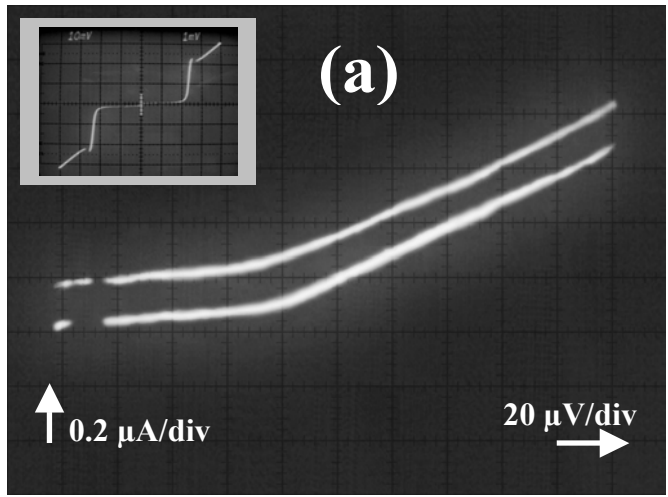

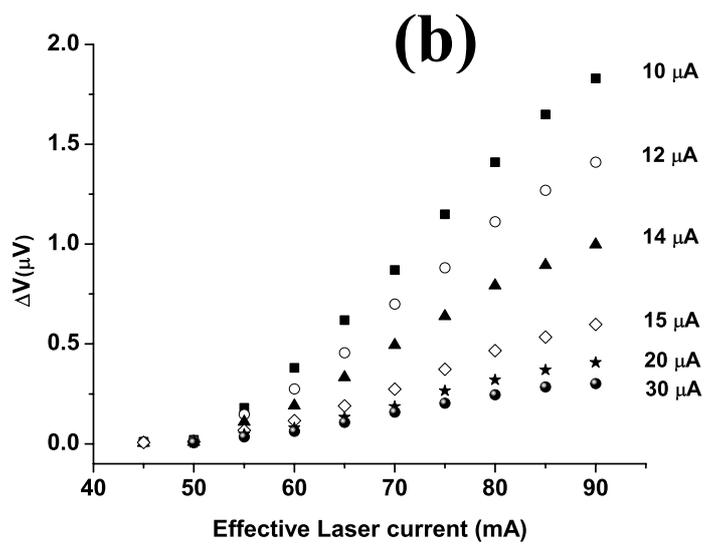

F.Stella et al., Fig.3